\documentclass[10pt,EJP,final,a4paper]{iopart}
\usepackage{newpxtext,newpxmath}
\usepackage{soul}
\usepackage{sfmath}
\usepackage{amsmath}
\usepackage{cite}
\usepackage{graphics}
\usepackage{bbm}
\usepackage[utf8]{inputenc}
\newcommand{\op}{\boldsymbol}
\newcommand{\U}{\uparrow}
\newcommand{\D}{\downarrow}
\newcommand{\LT}{+}
\newcommand{\RT}{-}
\usepackage[bookmarks=true,bookmarksnumbered=false,bookmarksopen=false, breaklinks=false,pdfborder={0 0 0},backref=false,colorlinks=true,citecolor=red,linkcolor=blue,urlcolor=blue]{hyperref}

\renewcommand\familydefault{\sfdefault} 
\begin{document}
\noindent\textbf{\textcolor{red}{IOP}} Publishing \hfill \textcolor{red}{European Journal of Physics}
\vspace{0mm}
\hrule
\vspace{2mm}

\noindent\EJP{\textbf{41}, 055403 (2020) \hfill \href{https://doi.org/10.1088/1361-6404/ab923e}{https://doi.org/10.1088/1361-6404/ab923e}}
\renewcommand{\familydefault}{phv}
\sffamily

\title{Demystifying the Delayed-Choice Quantum Eraser}
\author{Tabish Qureshi \href{https://orcid.org/0000-0002-8452-1078}{\resizebox{10px}{!}{\includegraphics{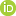}}}}
\address{Centre for Theoretical Physics, Jamia Millia Islamia, New Delhi,
India.}
\ead{\href{tabish@ctp-jamia.res.in}{tabish@ctp-jamia.res.in}}


\begin{abstract}
The delayed-choice quantum eraser has long been a subject of controversy,
and has been looked at as being incomprehensible to having retro-causal effect
in time. Here the delayed-choice quantum eraser is theoretically analyzed using
standard quantum mechanics. Employing Mach-Zehnder interferometer, instead
of a conventional two-slit interference, brings in surprising clarity.
Some common mistakes in interpreting
the experiment are pointed out. It is demonstrated that
in the delayed mode there is no which-way information present
after the particle is registered on the screen or the final detectors,
contrary to popular belief. However, it is
shown that another kind of path information is present even after the particle
is registered in the final detectors. The registered particle can be used
to predict the results of certain yet to be made measurements on the which-way
detector. This novel correlation can be tested in a careful experiment.
It is consequently argued that there is no big mystery in the experiment,
and no retro-causal effect whatsoever.
\end{abstract}


\section{Introduction}

\begin{figure}
\centerline{\resizebox{8.5cm}{!}{\includegraphics{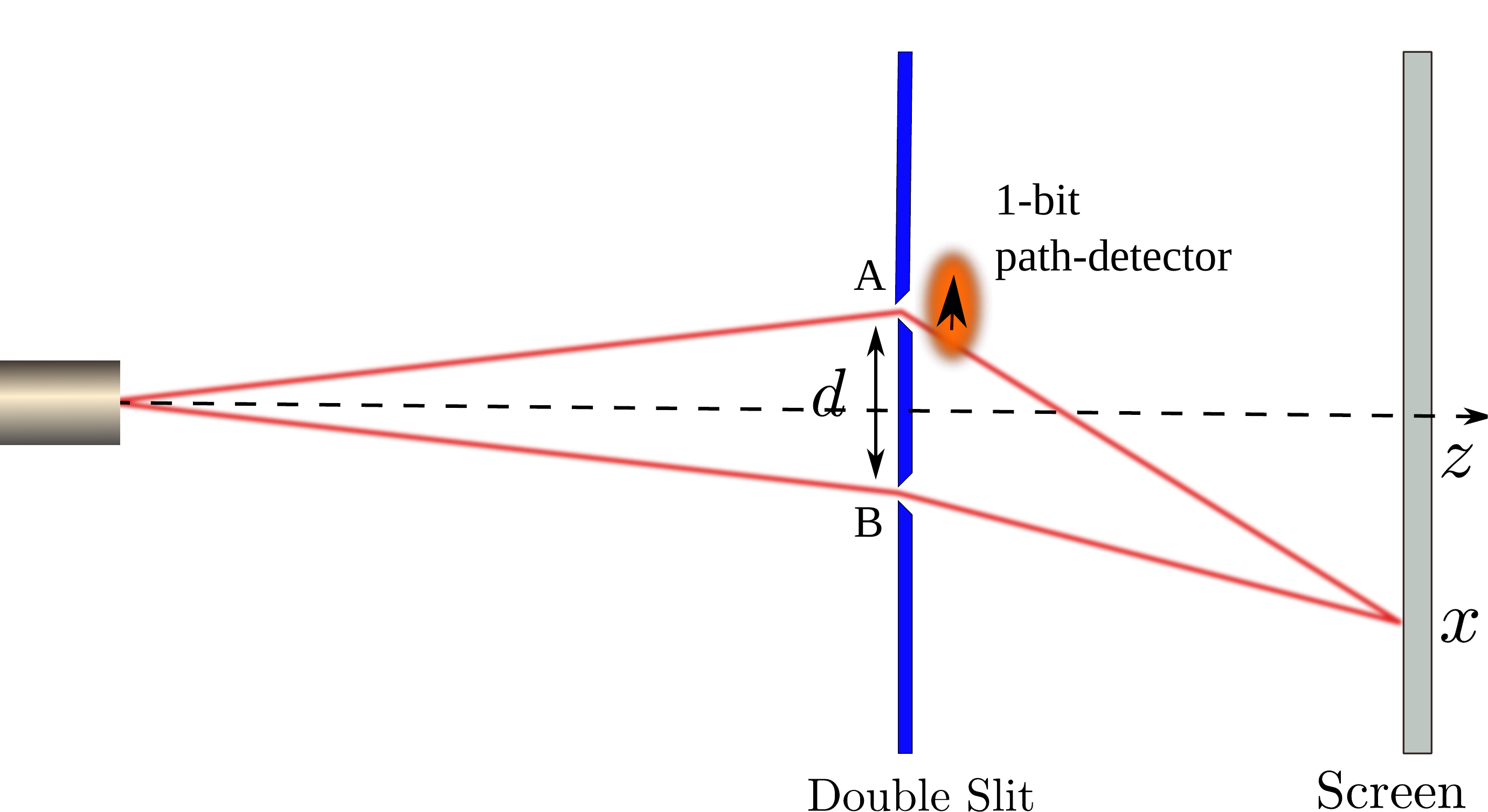}}}
\caption{Schematic diagram of a two-slit interference experiment in the
presence of a 1-bit which-way detector.  }
\label{eraser}
\end{figure}

Wave-particle duality, as it is understood today, is a concept that is
grounded in the principle of complementarity that Niels Bohr
formulated \cite{bohr}. Quantum objects, which we refer to as quantons,
can exhibit wave properties, akin to being "spread out", or particle
properties, akin to being localized. The two-slit interference experiment
has become a testbed for probing these and related issues \cite{einstein}.
In an oft-considered thought experiment, there is a 1-bit (two state)
quantum path detector sitting in the path of a quanton passing through a
double-slit (see FIG. \ref{eraser}). The two states of this "which-way"
detector are correlated with the two paths of the quanton. Reading the
state of the which-way detector can provide information regarding which slit
the quanton passed through. An interesting idea was advanced by
Jaynes \cite{jaynes}, according to which one may choose to look at such
states of the which-way detector which do not distinguish between the two
paths of the quanton, thus \emph{erasing} the which-way information.
This may enable bringing back the interference. 
Scully and Drühl \cite{druhl} also formulated such an idea and coined the
term "quantum eraser". Going further, they proposed that in a modified
experiment, one can choose to delay the erasing of the which-way information
until after the quanton is registered on the screen. This "delayed choice
quantum eraser", they showed, would also bring back interference.
The delayed-choice quantum eraser experiment led to a lively debate which
continues to this day \cite{esw,mohrhoff,aharonov,hiley,ellerman,taming,kastner}.
Lot of confusion prevailed over this proposed experiment, as to whether
it implies making the quanton behave like a wave or a particle, much 
after it has been registered on the screen. This apparent "retro-causality"
is still a subject of discussion \cite{ellerman,taming,kastner}.

Quantum eraser has now been experimentally realized by various people using
photons \cite{ma,mandel,chiao, zeilinger,kim-shih,walborn,kim,andersen,
scarcelli,neves,schneider}.
There have been some other proposals using neutral kaons \cite{bramon},
using a modified Stern-Gerlach steup \cite{zini,barney}, and also using atoms in
an optical Strern-Gerlach model \cite{chianello}. The idea of quantum
eraser has also been generalized to three-path interference \cite{3eraser}.

Here we take a fresh look at the delayed choice quantum eraser and analyze
various issues which have been under debate.

\section{Two-Slit Interference and Quantum Eraser}

In the following we briefly explain the basic idea behind quantum eraser.
Consider a quanton going through a double-slit, and 
let $|\psi\rangle$ be the state of the quanton when it emerges from
the double-slit:
\begin{equation}
|\psi\rangle = \tfrac{1}{\sqrt{2}} \left[ |\psi_1\rangle + |\psi_2\rangle \right],
\label{ent0}
\end{equation}
where are $\psi_1,\psi_2$ are states localized at the location of 
slits 1 and 2, respectively.
The states $\psi_1,\psi_2$ are orthogonal because of their spatial separation.
The quanton travels to the screen and the probability of it landing at a
position $x$ is given by 
\begin{eqnarray}
|\langle x|\psi(t)|^2 &=& \tfrac{1}{2} \left[ |\psi_1(x,t)|^2 + \psi_2(x,t)|^2
\right.\nonumber\\
&&\left. + \psi^*_1(x,t)\psi_2(x,t) + \psi^*_2(x,t)\psi_1(x,t) \right],
\end{eqnarray}
where the last two term represent interference. In the subsequent discussion
we will drop the label $t$, and will just assume the state on the screen to
be the time-evolved state.

The age-old question is, which slit did the quanton go through? To address
this question, let us introduce a which-way detector at the double-slit,
as shown in FIG. \ref{eraser}. Although which-way detection can be implemented
in a variety of ways, we just consider a \emph{1-bit} detector, like a
quantum spin$-1/2$, without assuming a specific form of it.
The which-way detector gets entangled with the states of the two paths, and
the combined state of the quanton and which-way detector is given by
\begin{equation}
|\Psi\rangle = \tfrac{1}{\sqrt{2}} \left[|\psi_1\rangle|\U\rangle + |\psi_2\rangle|\D\rangle \right],
\label{ent1}
\end{equation}
where $|\U\rangle, |\D\rangle$ are certain orthonormal states of the 
which-way detector, like the eigenstates of the z-component of a spin$-1/2$.
We will assume that the state of the quanton at the screen and the which-way
detector continues to be given by (\ref{ent1}), while remembering that
$\psi_1(x), \psi_2(x)$ at the screen would be the time-evolved states.
Treating explicit time evolution of the states is not important for the
purpose here, as what matters is that the entanglement in (\ref{ent1}) is
retained.  For the which-way detector states, as they are like states of a
spin-1/2, it can be assumed that there is no "free" Hamiltonian, so the
states do not change with time. This also makes it straightforward to 
decide whether one wants to look at the quanton registering on the screen
before measuring the which-way detector state, or vice-versa. For example,
if one wants to make a which-way measurement before the quanton hits the
screen, one has to project the entangled state on a particular which-way
detector state, and look at what state of the quanton emerges. On the
other hand, if one wants to measure the which-way detector after the 
quanton hits the screen, one only has to project the entangled state on
a particular position eigenstate at the screen, and see what state of 
the which-way detector is left behind. In both the cases, the state of
the quanton will be assumed to be the state when the quanton has reached
the screen.  One can now evaluate the probability density of
the quanton falling on the screen at a position $x$, namely
$|\langle x|\Psi\rangle|^2$, as 
\begin{eqnarray}
|\langle x|\Psi\rangle|^2 &=& \tfrac{1}{2} \left[ |\psi_1(x)|^2 + |\psi_2(x)|^2 \right.\nonumber\\
&&\left. + \psi_1^*(x)\psi_2(x)\langle \U|\D\rangle + 
\psi_2^*(x)\psi_1(x)\langle \D|\U\rangle\right].
\end{eqnarray}
The cross terms in the above, which represent interference, have a factor
proportional to $|\langle \U|\D\rangle|$, which is equal to zero, thus
destroying the interference of the quanton. The standard quantum lore is that
since the which-way detector ``carries" the which-way information about the
quanton (by virtue of the entangled state), the interference is destroyed.

\begin{figure}
\centerline{\resizebox{8.5cm}{!}{\includegraphics{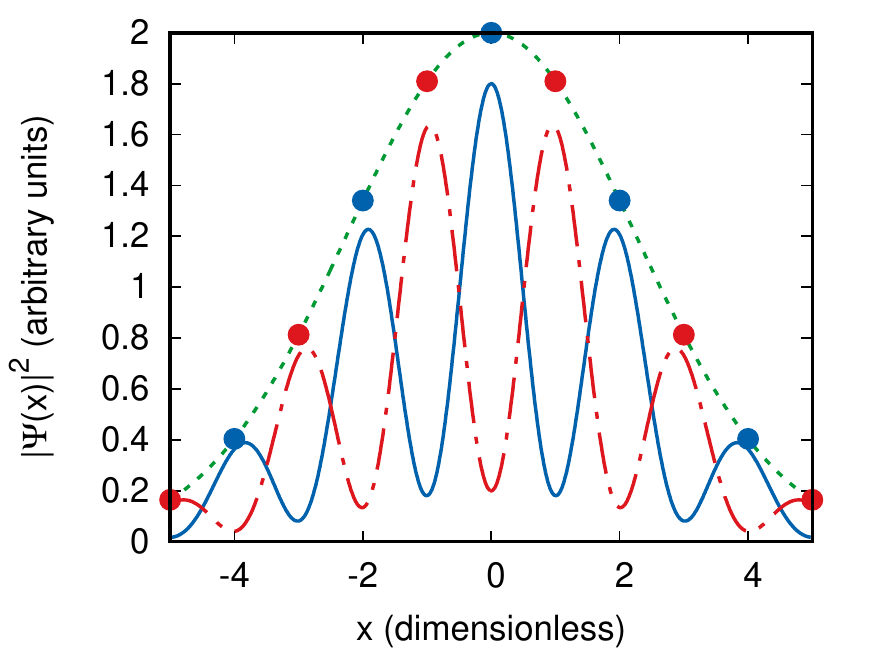}}}
\caption{Probability density of quantons falling on the screen, as a function
of position (green, dotted curve). They show no interference because of the
presence of the which-way detector. Probability density of quantons in 
coincidence with which-way detector landing in $|+\rangle$ state (blue, solid
curve). It shows an interference pattern, representing erasure of which-way
information.
Probability density of quantons in coincidence with the which-way detector
landing in $|-\rangle$ state (red, dashed curve). It too shows an interference
pattern, but phase-shifted. See text for the meaning of red and blue circles. }
\label{plot}
\end{figure}

A quantum eraser is introduced in the following manner.
If $|\U\rangle,|\D\rangle$ are orthonormal,
one can introduce another set of orthonormal states:
$|\pm\rangle =\tfrac{1}{\sqrt{2}}(|\U\rangle \pm |\D\rangle)$, which
are like eigenstates of the $x-$component of a spin$-1/2$.
The entangled state (\ref{ent1}) can then be written as
\begin{eqnarray}
|\Psi\rangle &=& \tfrac{1}{2} [ |\psi_1\rangle + |\psi_2\rangle ]|\LT\rangle 
 + \tfrac{1}{2} [ |\psi_1\rangle - |\psi_2\rangle ]|\RT\rangle.
\label{ent3}
\end{eqnarray}
It is obvious that (\ref{ent3}) shows no interference, as it is the same
state as (\ref{ent1}). However, if the quanton is detected in
coincidence with the state $|\LT\rangle$ of the which-way detector, it
shows an interference
which is exactly the same as that shown by (\ref{ent0}). Alternately,
if the quanton is detected in coincidence with the state $|\RT\rangle$,
it shows an interference which is slightly shifted. In this
sense, the states $|\pm\rangle$ may be called \emph{both-ways} states.
The two interferences may be represented as
\begin{eqnarray}
|\langle \LT|\Psi(x)\rangle|^2 &=& \tfrac{1}{4} \left[ |\psi_1(x)|^2 + |\psi_2(x)|^2 \right.\nonumber\\
&&\left. + \psi_1^*(x)\psi_2(x)+ \psi_2^*(x)\psi_1(x)\right],\nonumber\\
|\langle \RT|\Psi(x)\rangle|^2 &=& \tfrac{1}{4} \left[ |\psi_1(x)|^2 + |\psi_2(x)|^2 \right.\nonumber\\
&&\left. - \psi_1^*(x)\psi_2(x)- \psi_2^*(x)\psi_1(x)\right],
\label{interf}
\end{eqnarray}
The standard narrative says that which-way information, which was carried
by the  correlated state (\ref{ent1}), is \emph{erased} on obtaining a
both-ways state $|\LT\rangle =\tfrac{1}{\sqrt{2}}(|\U\rangle + |\D\rangle)$.
Because of the fact that coincident detection of the quanton with
$|{\pm}\rangle$ states brings back the interference, the process is
called quantum erasure \cite{druhl}.

Depending on which set of states of the which-way
detector one chooses to look at, one may choose to retain or erase the 
which-way information. If one measures the z-states of the which-way
detector before the quanton hits the screen, and finds (say) $|\D\rangle$,
one knows for sure that the quanton when through slit 2, and not through
slit 1. One can repeat this procedure for many quantons and for each of them
one knows which slit they went through. However, those quantons will not
form an interference pattern on the screen. Alternatively, one may decide
to measure the x-state of the which-way detector before the quanton hits
the screen. If one obtains $|\LT\rangle$, one knows that the state of the
quanton is $\tfrac{1}{\sqrt{2}} [ |\psi_1\rangle+ |\psi_2\rangle]$. This would mean that
the quanton went through both the slits, like a wave. If one obtains
$|\RT\rangle$, it implies that the state of the quanton is
$\tfrac{1}{\sqrt{2}} [ |\psi_1\rangle - |\psi_2\rangle]$. Here too, the quanton
went through both the slits, like a wave, but in a slightly different fashion.
Naturally, in these two cases, one does obtain an interference pattern.
Thus one can force the quanton to behave like a particle, or a wave, by
choosing which set of states of the which-way detector one measures.

A clarification may be in order here regarding a philosophical objection
that has been recently raised \cite{kastner}.
What do we mean when we say that the which-way information is \emph{carried}
about the quanton? We simply mean that the entangled
state (\ref{ent1}) has the potential to yield which-way information, through
an appropriate measurement. It is not implied that the quanton is actually
going through the path (say) $|\psi_1\rangle$ and the which-way detector
is $|\U\rangle$. In fact, one may equally well say that the entangled
state carries information about which of the two states,
$\tfrac{1}{\sqrt{2}} [ |\psi_1\rangle \pm |\psi_2\rangle]$ the quanton
may be found in. This again means that the correlated state has the 
potential to yield information on which of the two states
$\tfrac{1}{\sqrt{2}} [ |\psi_1\rangle \pm |\psi_2\rangle]$ the quanton
will be found in, if an appropriate measurement is done on the which-way
detector.

An apparently perplexing situation arises if one observes the which-way
detector much after the quanton has been registered on the screen. One may
still (rather naively) try to correlate the measurement results of the which-way
detector (in which ever basis) with the detection of the quanton on the screen.
It should be noted that the probabilities given by (\ref{interf})
are independent of whether one looks at the which-way detector before
or after the quanton hits the screen \cite{taming}. So, one would still
see no interference in coincidence with the states$|\U\rangle, |\D\rangle$,
but will see a ``recovered" interference in coincidence with either
$|\RT\rangle$ or $|\LT\rangle$, separately.
If one continues to use the logic of the preceding discussion, it
appears to imply that one can force the quanton to behave
like a particle or a wave, much after it has been registered on the screen.
This inference has perplexed people and led many to debate if quantum
mechanics allows one to have backward in time influence.

\section{Understanding quantum correlations}
\label{qcorr}

Let us first clearly understand the basis on which we infer the path followed
by the quanton from the state of the which-way detector. The inference is
a result of quantum entanglement and the resulting correlation between
the two. For simplicity, consider two spin$-1/2$ particles 1 and 2, in an
entangled state
\begin{equation}
|\phi\rangle = \tfrac{1}{\sqrt{2}}[|\U\rangle_1|\U\rangle_2 + |\D\rangle_1|\D\rangle_2],
\label{spinz}
\end{equation}
where labels 1,2 refer to the two particles, and states
$|\U\rangle_i$, $|\D\rangle_i$ denote the eigenstates of the z-component of
the spins. The same state can also be written as
\begin{equation}
|\phi\rangle = \tfrac{1}{\sqrt{2}}[|\LT\rangle_1|\LT\rangle_2 + |\RT\rangle_1|\RT\rangle_2],
\label{spinx}
\end{equation}
where the state $|\LT\rangle_i, |\RT\rangle_i$ denote the eigenstates of the
x-component of the spins. Now if one measures the z-component of spin 1,
and suppose finds it in the state $|\U\rangle_1$, one immediately
knows that the state of spin 2 is $|\U\rangle_2$, because the measurement
reduces the state $|\phi\rangle$ given by (\ref{spinz}) to
$_1\langle \U|\phi\rangle = |\U\rangle_2$. On the other hand, if one measured
the x-component of the spin of particle 1, and found (say) $|\RT\rangle_1$,
one would immediately know the state of particle 2 to be $|\RT\rangle_2$.
This is because the measurement reduces the state $|\phi\rangle$ given by
(\ref{spinx}) to $_1\langle \RT|\phi\rangle = |\RT\rangle_2$. 
So, because of the entangled state, there is a correlation between the
z-components of the two spins. For the same reason there is a correlation
between the x-components of the two spins. However, there is no correlation
between (say) the z-component of spin of particle 1 and the x-component of
spin of particle 2. This can be simply verified as follows. Measurement of
z-component of spin 1 and obtaining $|\D\rangle_1$ leads to
$_1\langle \D|\phi\rangle = |\D\rangle_2 = \tfrac{1}{\sqrt{2}}
(|\LT\rangle_2 - |\RT\rangle_2)$. If now one measures x-component of the spin
of particle 2 on this reduced state, one is equally likely to get 
$|\LT\rangle_2$ or $|\RT\rangle_2$. 

Now, the crucial point is the following. Since one knows that measuring
z-component of spin 1 will tell one about the z-component of spin-2, one
might be tempted to make it a always-holds-true rule. Now one first 
measures the x-component of spin 2, and finds (say) $|\LT\rangle$. Then
one decides to ask, what was the z-component of spin 2 before one measured its
x-component. One may naively use the above-mentioned always-holds-true rule,
and measure the z-component of spin 1, to find (say) $|\U\rangle_1$. One might
now claim, aha! this means that the z-component of spin 2, before
one measured its x-component, was $|\U\rangle_2$. But that is wrong, simply
because the correlation between the z-components of the two spins is 
based on the entangled state (\ref{spinz}), but this entangled state is 
already destroyed when one measured the x-component of spin 2.
This example will now help in identifying where the flaw in the delayed-choice
argument lies. 

Suppose spin 2 plays the role of the which-way detector in the quantum
eraser experiment, and spin 1 plays the role of possible paths of the
quanton in the following way:\\ $|\U\rangle_1 \rightarrow |\psi_1\rangle$,
$|\D\rangle_1 \rightarrow |\psi_2\rangle$, and
$|\pm\rangle_1 \rightarrow \tfrac{1}{\sqrt{2}}[|\psi_1\rangle\pm|\psi_2\rangle]$.
Just as for the the case of two spins, there is a correlation between 
the states $|\U\rangle,|\D\rangle$ and $|\psi_1\rangle,|\psi_2\rangle$.
Also, there is a correlation between $|\LT\rangle,|\RT\rangle$ and
$\tfrac{1}{\sqrt{2}}[|\psi_1\rangle\pm|\psi_2\rangle]$. And that is by virtue
of the entangled state given by (\ref{ent1}) and (\ref{ent3}).
But now suppose that the quanton registers on the screen at a position $x_0$.
The entangled state (\ref{ent1}) gets reduced to
\begin{equation}
\langle x_0|\Psi\rangle = \tfrac{1}{\sqrt{2}} \left[ \langle x_0|\psi_1\rangle|\U\rangle + \langle x_0|\psi_2\rangle|\D\rangle \right].
\label{ent1r}
\end{equation}
Since the entangled state is gone, one cannot use (\ref{ent1r}) to 
measure the z-state of the which-way detector and then infer from the result
as to whether the state of the quanton was $|\psi_1\rangle$ or $|\psi_2\rangle$
before it landed at position $x_0$. This point will be elaborated upon in the
next section. The analogy between a quantum eraser setup and entangled spins
is known \cite{esw}. In fact, Kastner has used a similar argument,
by identifying the $\tfrac{1}{\sqrt{2}}[|\psi_1\rangle\pm|\psi_2\rangle]$
states with the x-basis of a spin-1/2, to emphasize that once the quanton
registers on the screen, it yields no information regarding
the path followed by the quanton \cite{kastner,kastnerbook1,kastnerbook2}.

\begin{figure}
\centerline{\resizebox{8.5cm}{!}{\includegraphics{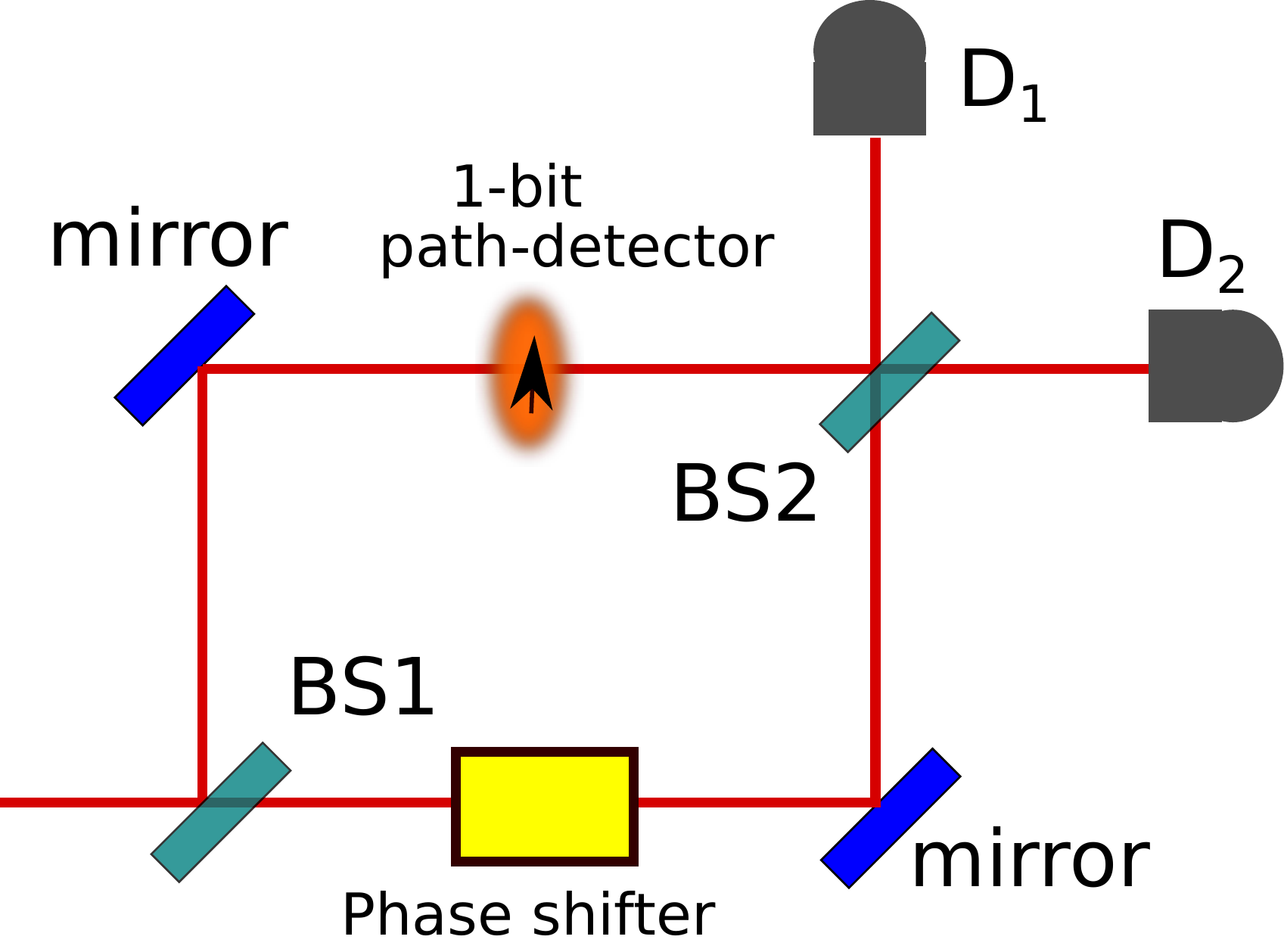}}}
\caption{A schematic diagram of a quantum eraser setup using a Mach-Zehnder
interferometer. There is a 1-bit quantum which-way detector in the path
of the quanton. }
\label{mz}
\end{figure}

\section{Quantum eraser using Mach-Zehnder interferometer}
\label{qemz}

Although the analogy between the entangled spins and the quantum eraser setup
is apparent, one may not be fully convinced because the quanton
involves a continuous variable, the position, and is not like a spin$-1/2$.
In order to make the analogy really one to one, we consider a 
Mach-Zehnder interferometer, with a 1-bit which-way detector, as shown in
FIG. \ref{mz} (see also Ref. \cite{ferrari}). A Mach-Zehnder interferometer can be analyzed using quantum
mechanics in the following way \cite{scarani}.
An incoming quanton in the state $|S\rangle$, gets split by the first 
beam-splitter BS1 into a spatially 
separated superposition $\tfrac{1}{\sqrt{2}}(|T\rangle-|R\rangle)$,
where $-$ sign represents a $\pi$ phase shift due to reflection.
The two component evolve, after reflecting from the mirrors and passing
through the beam-splitter BS2, to the final state at the detectors as follows:
$|T\rangle\rightarrow\tfrac{1}{\sqrt{2}}(-|D_1\rangle-|D_2\rangle)$ and
$|R\rangle\rightarrow\tfrac{1}{\sqrt{2}}(|D_1\rangle-|D_2\rangle)$, where
$|D_1\rangle, |D_2\rangle$, are the states at the detectors $D_1$ and $D_2$,
respectively.  To make the initial state appear the same as (\ref{ent0}),
one can redefine the states $|T\rangle, |R\rangle$ as $|\psi_1\rangle, |\psi_2\rangle$
by absorbing certain phase factors, to
write the initial state after the first beam splitter as
\begin{equation}
|\psi_I\rangle = \tfrac{1}{\sqrt{2}}(|\psi_1\rangle+|\psi_2\rangle).
\label{ent0mz}
\end{equation}
After the second beam-splitter, the two components
$|\psi_1\rangle, |\psi_2\rangle$ evolve to
\begin{eqnarray}
\op{U}_{BS2}|\psi_1\rangle &=& \tfrac{1}{\sqrt{2}}(|D_1\rangle+|D_2\rangle) \nonumber\\
\op{U}_{BS2}|\psi_2\rangle &=& \tfrac{1}{\sqrt{2}}(|D_1\rangle-|D_2\rangle) ,
\label{ubs2}
\end{eqnarray}
where $\op{U}_{BS2}$ represents the unitary evolution due to the mirrors
and the second beam-splitter BS2, and $|D_1\rangle, |D_2\rangle$ are the
states at the detectors $D_1, D_2$, respectively.
It might be interesting to see what the state (\ref{ent0mz}) will result in,
at the final detectors. The final state, just before the quanton hits the
detectors, is given by
\begin{eqnarray}
|\psi_f\rangle = \op{U}_{BS2}|\psi_I\rangle &=& \tfrac{1}{\sqrt{2}}
(\op{U}_{BS2}|\psi_1\rangle+\op{U}_{BS2}|\psi_2\rangle).
\label{mzf0}
\end{eqnarray}
Using (\ref{ubs2}) it is straightforward to see that the probability of the
quanton to end up at detector $D_1$ is $|\langle D_1|\psi_f\rangle|^2 = 1$
and the probability for it to end up at detector $D_2$ is
$|\langle D_2|\psi_f\rangle|^2 = 0$. This represents interference, as 
detector $D_1$ registers a bright fringe (all quantons landing there), whereas
detector $D_2$ registers a dark fringe (no quanton landing there).

Next we consider the effect of introducing a 1-bit which-way detector in
the path of the quanton. The combined state of the quanton and which-way
detector, after it passes through the first beam-splitter, and interacts
with the which-way detector, is given by
\begin{equation}
|\Psi_I\rangle = \tfrac{1}{\sqrt{2}} \left[ |\psi_1\rangle|\U\rangle + |\psi_2\rangle|\D\rangle \right].
\label{mzi}
\end{equation}
After passing through BS2, the combined state is
$|\Psi_F\rangle = \op{U}_{BS2}|\Psi_I\rangle$ :
\begin{equation}
|\Psi_F\rangle = \tfrac{1}{{2}} \left[(|D_1\rangle+|D_2\rangle)|\U\rangle
+ (|D_1\rangle-|D_2\rangle)|\D\rangle \right].
\label{mzz}
\end{equation}
If one finds which-way detector in the state $|\U\rangle$, the quanton state
is $\tfrac{1}{\sqrt{2}} (|D_1\rangle + |D_2\rangle)$, which means $D_1$
and $D_2$ are equally likely to click. If one finds which-way detector in the
state $|\D\rangle$, the quanton state
is $\tfrac{1}{\sqrt{2}} (|D_1\rangle - |D_2\rangle)$, which again
means $D_1$ and $D_2$ are equally likely to click. Thus, there is no 'dark fringe',
and hence no interference.

Next we look at the case where $D_1$ or $D_2$ register the 
quanton first, and much later one chooses to look at a particular basis of the
which-way detector. If (say) $D_1$ clicks (quanton state is $|D_1\rangle$),
eqn (\ref{mzz}) tells us that the 
which-way detector is now in the state $\tfrac{1}{\sqrt{2}}
(|\U\rangle + |\D\rangle)$. Measuring the which-way detector in the 
z-basis, one is equally likely to find $|\U\rangle$ or $|\D\rangle)$, 
which yields no which-way information. Thus we see that as soon as the
quanton registers at a detector, all which-way information is lost.
This is not surprising because eqn. (\ref{mzz}) says that $|D_1\rangle,
|D_2\rangle$ are not correlated to $|\U\rangle, |\D\rangle)$, rather to
$\tfrac{1}{\sqrt{2}} (|\U\rangle + |\D\rangle)$ and 
$\tfrac{1}{\sqrt{2}} (|\U\rangle - |\D\rangle)$. This shows that registering
the quanton at $D_1$ or $D_2$ destroys the which-way information. In the
delayed mode, even though there is no interference, clicks of $D_1$ and $D_2$
do not yield
any which-way information. This aspect has not been understood in any previous 
analysis, and has been a source of confusion in interpreting delayed-choice
experiments.

Another point to notice is that if one chooses to forget about the which-way
detector completely, and only calculates the probability of the quanton to
hit $D_1$ or $D_2$, one finds that 
$|\langle D_1|\psi_F\rangle|^2 = |\langle D_2|\psi_F\rangle|^2 = 1/2$.
This implies no interference as $D_1$ and $D_2$ will register equal number
of quantons. Although the which-way information is lost as soon as the 
quanton hits the detectors, the interference is lost too. Contrast this
with the case where the quanton is not entangled with any which-way detector,
namely, eqn. (\ref{mzf0}). In that case, the probability of the quanton
to hit $D_1$ is 1, and that to hit $D_2$ is 0. Entanglement with the 
which-way detector is enough to destroy the interference.

Interference is recovered in the usual quantum eraser experiments, by
correlating the clicks in the detectors $D_1$, $D_2$, with the x-states
$|+\rangle, |-\rangle$ of the which-way detector. Eqn. (\ref{mzz}), when
written in terms of these states, has the following form
\begin{equation}
|\Psi_F\rangle = \tfrac{1}{{\sqrt{2}}} \left[|D_1\rangle|+\rangle
+ |D_2\rangle|-\rangle \right].
\label{mzx}
\end{equation}
If the which-way detector is first looked at in the x-basis, and one finds
(say) $|+\rangle$, it means that $D_1$ will detect the quanton, and not 
$D_2$. When this happens for many quantons, it implies interference where the
'bright fringe' is at $D_1$ and the 'dark fringe' at $D_2$. Note that by
reading the which-way detector in the x-basis $|\pm\rangle$, we \emph{choose
to erase} the which-way information. This is in agreement with Bohr's 
complementarity principle because when interference is observed, there is
no which-way information.  Similarly, if the
which-way detector is found in the state $|-\rangle$, it implies that $D_2$
will definitely detect the quanton, and not $D_1$. Many such quantons
constitute interference where the 'bright fringe' is now at $D_2$, instead of
$D_1$. This interference is complementary to the one seen in correlation with
$|+\rangle$, and taken together the two imply no interference.

If this part of the experiment is carried out in the delayed mode, the
scenario becomes more interesting. It was already shown in the preceding
discussion that in the delayed mode, as soon as the quanton registers at
$D_1$ or $D_2$, the which-way information is erased. However, that is not
enough to get back interference. In addition, one has to measure the
which-way detector, and correlate each detected quanton
with $|+\rangle$ or $|-\rangle$, to get two interferences which are 'shifted'
with respect to each other. Bright fringe of one is the dark fringe of the
other. 

However, there is another aspect of it which has not been recognized in the
earlier studies of delayed-choice experiments. Notice that (\ref{mzx}) implies
that looking at which detector the quanton has landed in, one can now
\emph{predict} which of the two which-way detector states,
$|+\rangle$ or $|-\rangle$,  will be surely obtained in a measurement. So,
even though the which-way information is erased after the quanton is registered
in a detector, the quanton retains another kind of information about the
which-way detector. This can easily be tested in correlated measurements in
a delayed-choice quantum eraser experiment. So, it is not true that in the
delayed mode, the 
which-way information is erased only after the which-way detector is
looked at in the x-basis $|\pm\rangle$, as is widely believed. Not only does
the quanton registering at a detector erases the which-way information, it
additionally retains information about precisely how it is erased,
as $\tfrac{1}{\sqrt{2}} (|\U\rangle + |\D\rangle)$ or as
$\tfrac{1}{\sqrt{2}} (|\U\rangle - |\D\rangle)$. 

To summarize the conclusions of this section, when the which-way detector
is measured before the quanton hits the final detectors, one can \emph{choose}
to either \emph{obtain} the which-way information by reading z-basis states
$|\U\rangle, |\D\rangle$ or \emph{erase} it by reading x-basis
states $|\pm\rangle$.
If the quanton hits the final detectors before the which-way detector
is measured, the which-way information is erased, always. Reading out z-basis
states $|\U\rangle, |\D\rangle$ does not yield any which-way information. 
However, reading out x-basis states $|\pm\rangle$ states allows one to recover
two complementary interference patterns. More interestingly, every registered
quanton can be used to \emph{predict}, which of the states $|+\rangle,|-\rangle$
will be obtained if one measures the which-way detector after a delay! 

\section{Discussion}

One can now make a comparison of the quantum eraser experiment using
Mach-Zehnder setup with the two entangled spins considered in
section \ref{qcorr}. Let us first write (\ref{mzx}) in a slightly
different form
\begin{equation}
|\Psi_F\rangle = \op{U}_{BS2}|\Psi_I\rangle = \tfrac{1}{{\sqrt{2}}} \left[(|D_1\rangle|+\rangle
+ |D_2\rangle)|-\rangle \right].
\end{equation}
Using (\ref{ubs2}), one can write the above as
\begin{eqnarray}
\op{U}_{BS2}|\Psi_I\rangle &=& \tfrac{1}{{\sqrt{2}}}
\op{U}_{BS2}\left[\tfrac{|\psi_1\rangle+|\psi_2\rangle}{\sqrt{2}}|+\rangle
+ \tfrac{|\psi_1\rangle-|\psi_2\rangle}{\sqrt{2}}|-\rangle \right]\nonumber\\
|\Psi_I\rangle &=& \tfrac{1}{{\sqrt{2}}}
\left[\tfrac{|\psi_1\rangle+|\psi_2\rangle}{\sqrt{2}}|+\rangle
+ \tfrac{|\psi_1\rangle-|\psi_2\rangle}{\sqrt{2}}|-\rangle \right],
\label{mzx1}
\end{eqnarray}
which means that the correlation between the clicks of $D_1,D_2$ and
the state $|\pm\rangle$, is coming from the correlation between
$\tfrac{|\psi_1\rangle\pm|\psi_2\rangle}{\sqrt{2}}$ and $|\pm\rangle$,
contained in the initial entangled state. On the other hand, 
$|\psi_1\rangle, |\psi_2\rangle$ are correlated with $|\U\rangle,|\D\rangle$,
by virtue of (\ref{mzi}).
Now compare (\ref{mzi}) and (\ref{mzx1}) with (\ref{spinz}) and (\ref{spinx}).
Quanton states $|\psi_1\rangle, |\psi_2\rangle$ play the role of 
$|\U\rangle_1,|\D\rangle_1$, whereas 
$\tfrac{|\psi_1\rangle\pm|\psi_2\rangle}{\sqrt{2}}$ play the role of
$|\pm\rangle_1$. The which-way detector can be assumed to play the role of
Spin 2.
For the entangled spins, measuring z-component of spin 2 gives information
about the possible outcome of the z-component of spin 1, and vice-versa.
Measuring x-component of spin 2 gives information about the possible
outcome of the x-component of spin 1, and vice-versa. If x-component of
spin 1 is measured, it destroys all potential information about the
result of a future measurement of z-component of spin 2.
Exactly in the same way, as soon as $D_1$ clicks, the quanton state
is $\tfrac{|\psi_1\rangle+|\psi_2\rangle}{\sqrt{2}}$, which implies
that the both-ways state, when measured, will be $|+\rangle$. However,
now one cannot get any information regarding the which-way states
$|\U\rangle,|\D\rangle$, and hence no which-way information about the
quanton.

One might wonder why it was not realized earlier that in the delayed
mode the quanton registered on the screen can give information about 
which of the states $|\pm\rangle$ one would obtain if which-way detector
is measured. Probable reason for it is that most analyses use two-slit
interference instead of the Mach-Zehnder setup. While Mach-Zehnder
interferometer has only two output states, a double-slit interference 
consists of a multitude of position states of the quanton. Although
interference is most commonly studied in a two-slit experiment,
the interference is not very 'clean' in the sense that the bright and
dark fringes are not well separated. In Mach-Zehnder experiment, the
bright and dark fringes are well separated, and even registered on separate
detectors. However, it is indeed possible to guess the outcome of measurement
of which-way states $|\pm\rangle$, by looking at the quantons hitting the
screen in a two-slit interference experiment too. Typical interference patterns,
in a quantum-eraser two-slit experiment, are shown in FIG. \ref{plot}.
The red (solid) curve represents interference in coincidence with $|+\rangle$,
whereas the blue (dashed) curve represents interference in coincidence with $|-\rangle$.
Without any coincident detection, a quantum falling anywhere on the
screen could belong to either red curve or blue curve. However,
notice that the peaks of the red curve are located exactly at the minima
of the blue curve. Since one knows that it is a two-slit interference whose
parameters are known, one has information about the exact locations of the
maxima, minima of the \emph{would be} interference pattern. All the quantons
will fall on the green curve, which represents no interference. However,
if a quanton falls on the position of a maximum of the
red curve, it means it has zero probability to belong to the blue curve.
Such points are denoted in red in FIG. \ref{plot}.
So, it must actually belong to the red curve, and now one can predict that
a measurement on the which-way detector will definitely yield $|+\rangle$
state. One the other hand, if a quanton falls on the position of a maximum
of the blue curve, it means it has zero probability to belong to the red curve.
Such points are denoted in blue in FIG. \ref{plot}.
Now one can predict with certainty that the which-way detector will 
yield $|-\rangle$. This effect is novel, and can be easily tested in a
delayed-choice quantum eraser experiment. Needless to say, the conditions
for sharp interference should be there, for this to work.
The effect is more stark in the Mach-Zehnder implementation of the delayed
choice quantum eraser. As soon as the quanton is detected at $D_1$ ($D_2$),
the state of the which-way detector changes to $|+\rangle$ ($|-\rangle$).
It is trivial to see that if one tries to measure the z-states of the which
way detector, one \emph{does} get either $|\U\rangle$ or $|\D\rangle$,
but that \emph{does not} imply any which-way information.

What mental picture of the quanton
traversing the two Mach-Zehnder paths should one construct, one might ask.
The mental picture consistent with the preceding analysis is that
if the which-way detector is not measured before the quanton registers at
the final detectors, the quanton does pass through both the paths, like
a wave, but 
the phase difference between the two paths is determined only when the
quanton ends up at $D_1$ or $D_2$. To fix the phase difference between the
two paths, one need not measure the x-states of the which-way detector.
Each click of $D_1$ or $D_2$ uniquely determines the corresponding x-basis
state of the which-way detector, and also the phase difference between the
two paths. As the phase difference varies between two values from quanton
to quanton, all of them taken together show no interference.
On the other hand, if the which-way
detector is measured, in the z-basis, \emph{before} the quanton hits the
detectors $D_1$, $D_2$, the mental picture one may construct is that the
quanton actually goes through only one of the two paths like a 
particle, unlike a wave. Measuring the z-state kills one of the paths
of the quanton, irrespective of how far away along the two paths the
quanton has traveled before that measurement is made.
So, the bottom line is that which-way
information about the quanton can only be obtained before the quanton
registers at $D_1$ or $D_2$.

\section{Conclusion}

In conclusion, we have theoretically analyzed the delayed choice quantum
eraser experiment in a two-slit interference setup and also using a
Mach-Zehnder setup. This is done by introducing a 1-bit quantum which-way
detector in the path of the quanton. We have first discussed the quantum
correlations arising
from entanglement, which form the basis on which one uses which-way 
detector to get which-way information about the quanton. If the which-way
detector is measured before the quanton is registered at the final
detectors (or the screen), one can \emph{choose} to either retrieve
which-way information about the quanton by reading the which-way detector
in the z-basis, or erase the which-way information by reading the 
which-way detector in the x-basis. In the latter case, two complementary
interferences can be recovered by correlating the detected quantons with
the x-basis states $|\pm\rangle$ of the which-way detector.

If the quanton is registered at the detectors (or the screen) before the
which-way detector is measured, the which-way information is erased, and
the state of the which-way detector is set by the process of registering
of the quanton.  This is proved
by the fact that final detectors $D_1, D_2$ can be used to \emph{predict}
which state of the x-basis of the which-way detector will emerge if a
measurement is made on it after a delay. For example, when quanton lands
at the detector $D_2$, according to the entangled state, the state of the
which-way detector changes to $|-\rangle$. Two interferences can again
be obtained by correlating with the x-basis states. Since the which-way
x-basis state is already decided once the quanton lands at $D_1$ or $D_2$, it
is obvious that if one chooses to measure z-basis instead, it is
not going to yield any which-way information. Not only that, by choosing
to look at z-basis states, one loses the opportunity to recover the
interference which is seen only in correlation with x-states. However,
the loss of interference here does not imply that there is any which-way
information present. It is not, contrary to popular belief.

In the two-slit implementation of quantum eraser, the quanton, in the 
delayed mode, landing on the screen cannot always predict the x-state of
the which-way detector simply because dark and bright fringes are not 
cleanly separated as in the Mach-Zehnder interferometer. However, there is
no conceptual difference between the two. The quanton landing on 
certain specific positions can indeed predict the x-state of the which-way
detector. This can be tested in a careful experiment. In the light of this
analysis, there is no mystery in the delayed choice quantum eraser experiment,
and no question of any retro-causality.

\end{document}